\documentclass[epj,twocolumn]{webofc}
\usepackage[varg]{txfonts}   
\woctitle{ISVHECRI 2016}
\begin{document}
\title{Primary gamma ray selection in a hybrid timing/imaging Cherenkov array}
%
%

\author{E.B. Postnikov\inst{1}\fnsep\thanks{Corresponding author, \email{evgeny.post@gmail.com}} \and
        A.A. Grinyuk\inst{2} \and
        L.A. Kuzmichev\inst{1} \and
        L.G. Sveshnikova\inst{1}
}

\institute{Lomonosov Moscow State University Skobeltsyn Institute of Nuclear Physics (MSU SINP), 1(2) Leninskie gory, GSP-1, 119991, Moscow, Russia
\and
           Joint Institute for Nuclear Research, 6 Joliot-Curie, 141980, Dubna, Moscow region, Russia
          }

\abstract{%
This work is a methodical study on hybrid reconstruction techniques for hybrid imaging/timing Cherenkov observations. This type of hybrid array is to be realized at the gamma-observatory TAIGA intended for very high energy gamma-ray astronomy (>30 TeV). It aims at combining the cost-effective timing-array technique with imaging telescopes. Hybrid operation of both of these techniques can lead to a relatively cheap way of development of a large area array. The joint approach of gamma event selection was investigated on both types of simulated data: the image parameters from the telescopes, and the shower parameters reconstructed from the timing array. The optimal set of imaging parameters and shower parameters to be combined is revealed. The cosmic ray background suppression factor depending on distance and energy is calculated. The optimal selection technique leads to cosmic ray background suppression of about 2 orders of magnitude on  distances up to 450~m for  energies greater than 50~TeV.}
\maketitle
\section{Introduction}
\label{intro}
\begin{figure}[ht]
\centering
\includegraphics[height=5cm,width=8cm,clip]{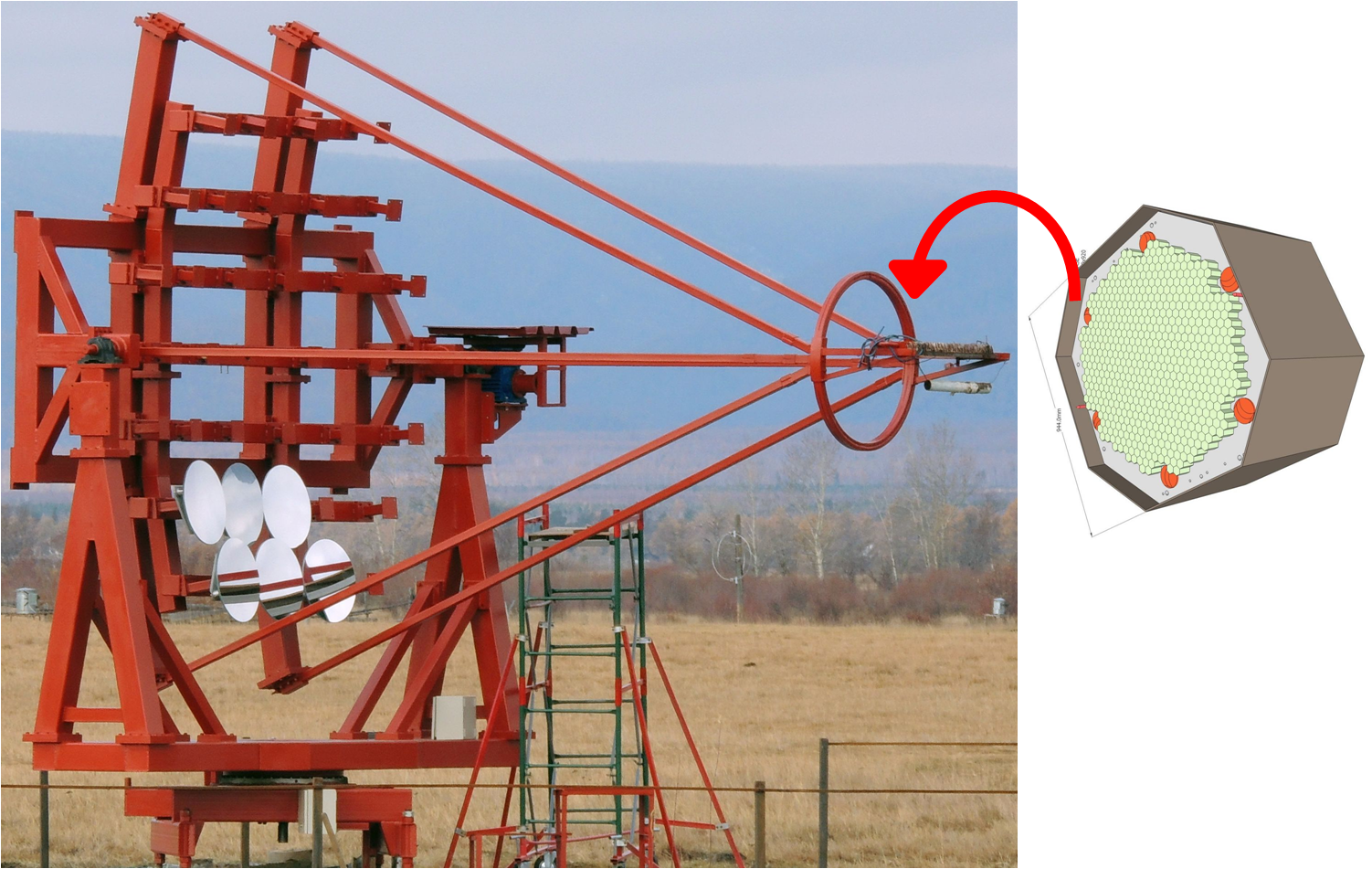}
\caption{An IACT in the deployment phase (left panel) and a camera (right panel).}
\label{fig-1}       
\end{figure}

Of several methods to detect primary gamma rays on the Earth's surface, an imaging technique is the most popular because of the  high efficiency of cosmic ray background reduction \cite{1}. Realization of this technique requires expensive and complicated installations called IACT (Imaging Air Cherenkov Telescopes) \cite{2}. They collect Cherenkov light from extensive air showers by a set of mirrors and reflect it onto an array of photomultipliers called a camera in such a way that a 2-dimensional image of the event is formed (Fig. \ref{fig-1}). The properties of an image are used to discriminate gamma ray induced showers against showers from the cosmic ray background.

The modern version of IACT implementation is stereoscopy \cite{3}, requiring a system of two or more imaging telescopes located at a distance of $\sim$100~m from each other \cite{4}. The reason for using  stereoscopy is the ability to determine the  core position of a gamma ray shower, as well as the shower arrival direction, because different telescopes detect images of the same shower at different angles.

Another technique to detect primary gamma rays is measuring the arrival time of the Cherenkov light front along with the local intensity. This method requires a set of wide angle timing optical stations and for the first time was implemented in such experiments as THEMISTOCLE \cite{5} and AIROBICC \cite{6}. That kind of technique is much more cost-effective, the installations have a wide angle of view, and arrays of detectors can easily be developed on a large area. A significant  disadvantage of this approach is a poor rate of cosmic ray background rejection. Some efforts \cite{7} have been made by the authors of the present work to assess the ability of gamma ray selection based on differences in the form of Cherenkov light lateral distribution. The research revealed that such a selection can only  be done  for events with a large number of triggered detectors, but anyway the selection quality is still much weaker than that of the IACT technique.  

All these reasons were taken into account when developing the gamma-ray observatory TAIGA (Tunka Advanced Instrument for cosmic ray physics and Gamma-ray Astronomy \cite{8}). TAIGA combines the hybrid detection of air shower Cherenkov radiation by the wide angle timing array TAIGA-HiSCORE and the narrow angle imaging telescopes TAIGA-IACT \cite{8}. At present, the first prototype of the telescope is being deployed in the Tunka valley (Fig. \ref{fig-1}). The telescope mirrors have an area of $\sim$10~m$^2$ and a focal length of 4.75~m, the camera consists of 560 photomultipliers with a total field of view 9.72$^\circ$ (0.36$^\circ$ per pixel).

Preliminary results of simulation confirm that the joint operation of the HiSCORE array and an IACT could be a very cost-effective way to increase the energy range of gamma-ray astronomy \cite{9}.

This work is a methodical study on hybrid reconstruction techniques to be used within the TAIGA experiment.

\section{Monte Carlo simulation}
\label{sec-2}

Simulated data of both types were used: the image parameters from the IACT, and the shower parameters reconstructed from the data obtained with the timing array. 

The IACT data was simulated in a few consecutive steps. First, the shower development in the air was simulated by CORSIKA \cite{10} to obtain the Cherenkov photon distribution on the observation level. Second, Cherenkov photons were traced through the optical system of the IACT. Third, conversion from photons to photoelectrons in every photomultiplier of a camera was carried out using wavelength-dependent atmospheric absorption and photo-cathode sensitivity. Then the signal was sorted into 40 ns time bins and a night sky background  added. Finally, a Cherenkov image was reproduced as a distribution of the number of photoelectrons in 560 photomultipliers. 

The timing array data provides the arrival direction (with an accuracy of 0.1$^\circ$--0.4$^\circ$ depending on energy), shower core position with   10 m accuracy, energy resolution with   $\sim$15$\%$ \cite{8} accuracy. As a first step the monoenergetic data banks with  primary particle energies from 20~TeV to 500~TeV were examined. 

Gamma ray showers direction was fixed at the point source, whereas the background (proton) shower arrival direction was randomized around the gamma ray shower direction within a cone of $0.4^\circ$ corresponding to the arrival direction resolution. This imitates the first selection step of showers in the cone of 0.4$^\circ$ based only on the timing array data.

\section{Methods}
\label{sec-3}

\subsection{Accuracy assessment}

Data analysis technique for the IACT is usually based on the "Hillas analysis" \cite{11}. For our task this method was modified. For each shower from the data bank various features of the image were calculated, and for each set of these features the selection quality factor, $Q$, was estimated as an accuracy indicator of our method.

The selection quality factor shows how much a significance, $S_0$,  of the statistical hypothesis that the events detected by the installation do not belong to the background, can be improved after the event selection:
\begin{equation}
S=S_0{\times}Q
\end{equation}
In the framework of Poisson statistics, the significance has a simple form:
\begin{equation}
S=N_{\gamma}/\sqrt{N_{bckgr}},
\end{equation}
and thereby the selection quality factor is:
\begin{equation}
Q=\epsilon_{\gamma}/\sqrt{\epsilon_{bckgr}},
\end{equation}
where $\epsilon_{\gamma}$ and $\epsilon_{bckgr}$ are the fractions of gamma ray events and background events respectively after the selection stage. Optimal algorithms (and optimal parameter sets) were found in a process of maximizing the Q factor value under the condition:
\begin{equation}
\epsilon_{\gamma}\geq0.5
\end{equation}
The fraction of background events after the selection corresponding to the $Q$ value is:
\begin{equation}
\epsilon_{bckgr}=(\epsilon_{\gamma}/Q)^2
\end{equation}

\subsection{Image cleaning}

For every event the night sky background light was simulated $N$ times to obtain more precise results, and the selection quality factor was averaged over all realizations. This multiple recalculation of background proved necessary because of strong fluctuations of the Q factor caused by random distortion of an image \cite{12} (distributions of $Q$ were plotted in \cite{12}). 

To perform image cleaning from random distortion, a few types of filtering were examined and the best algorithm as well as the optimal filtering parameters were chosen based on the $Q$ maximization criterion \cite{12}. Optimal filtering parameters in an image cleaning procedure are energy-dependent and distance-dependent, and their influence on $Q$ is sufficient \cite{12}. All the results below were obtained after  optimal image filtering.

\section{Results}
\label{sec-4}
\subsection{Image parameters}

The optimal set of parameters to be combined in a selection algorithm was revealed. 
The optimal imaging parameters from the IACT are:
\begin{itemize}
\item azwidth \cite{11}, or "azimuthal width", that is the "r.m.s. image width relative to an axis which joins the source to the centroid of the image" \cite{11}. The telescope is supposed to point at the gamma ray source, so the ``source'' on the image plane is simply the geometrical center of the camera;
\item core-azwidth, or "core-azimuthal width", suggested in the present article. This is the r.m.s. image width relative to an axis which joins the image centroid to the shower core position. 
\end{itemize}

\subsection{Optimal cuts}

\begin{figure}
\centering
\includegraphics[height=6.5cm,clip]{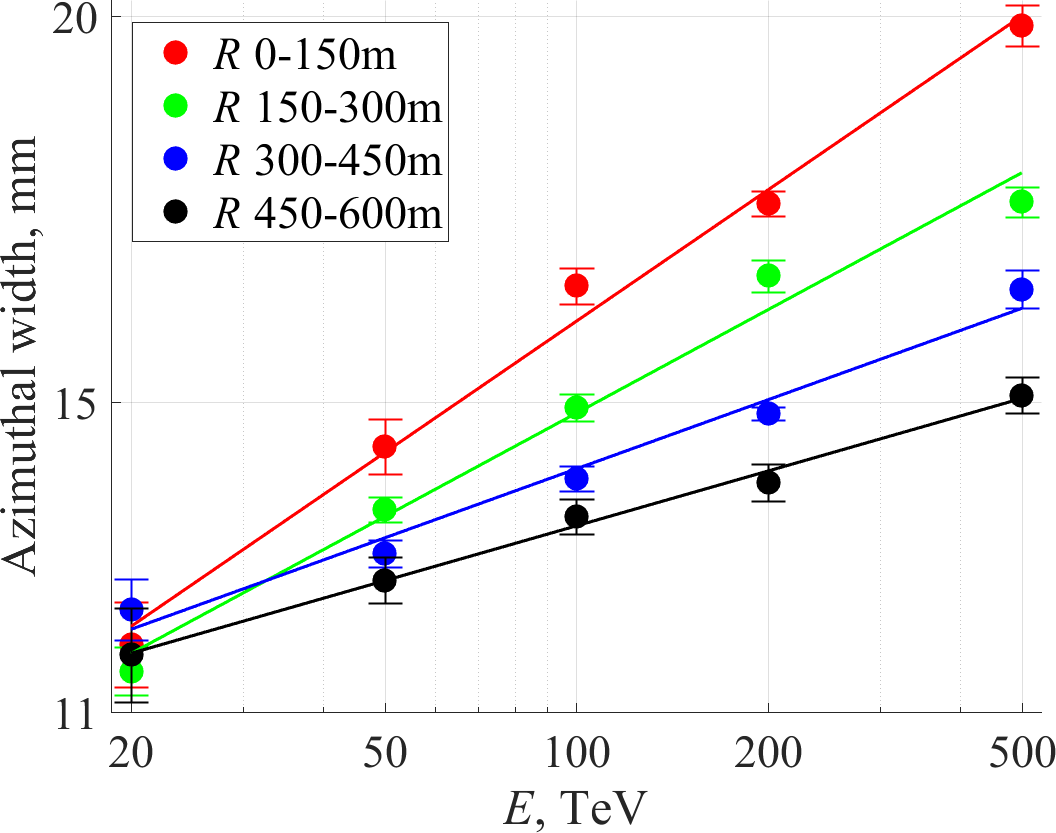}
\caption{Selection cut depending on energy and distance.}
\label{fig-2}       
\end{figure}

For every sample of events with fixed energy and interval of distance to the core position we have found the optimal cuts, corresponding to a maximal value of $Q(R,E)$. In Fig.~\ref{fig-2} we present the obtained optimal azwidth cut depending on energy and distance to the shower core.

While the optimal cuts depend on energy and distance, one can conclude that the parameters determined by the timing array (primary energy, shower core position, shower arrival direction) are very essential for gamma ray selection. 

An alternative way to take account of this dependence is  by normalizing the Hillas parameters (image width etc.) depending on shower parameters. The Hillas parameters after normalization are called ``scaled'' (``scaled width'' etc.) \cite{13}. They are used in stereoscopy (section \ref{intro}), where the core distance is obtained as an intersection of axes of multiple images. In a  previous analysis of TAIGA simulation we also followed this approach (''hybrid scaled width'' \cite{9}), however, in the present work the scaling procedure is applied to the selection cuts. An additional difference of the present approach is scaling the cuts depending on both the core distance and primary energy from the timing array of TAIGA. Our previous steps \cite{9} followed standard IACT ``scaled cuts'' analysis using total number of photoelectrons in an image (``image size'') instead of the energy.
  
\subsection{Optimal Q factor}

\begin{figure}
\centering
\includegraphics[height=6.5cm,clip]{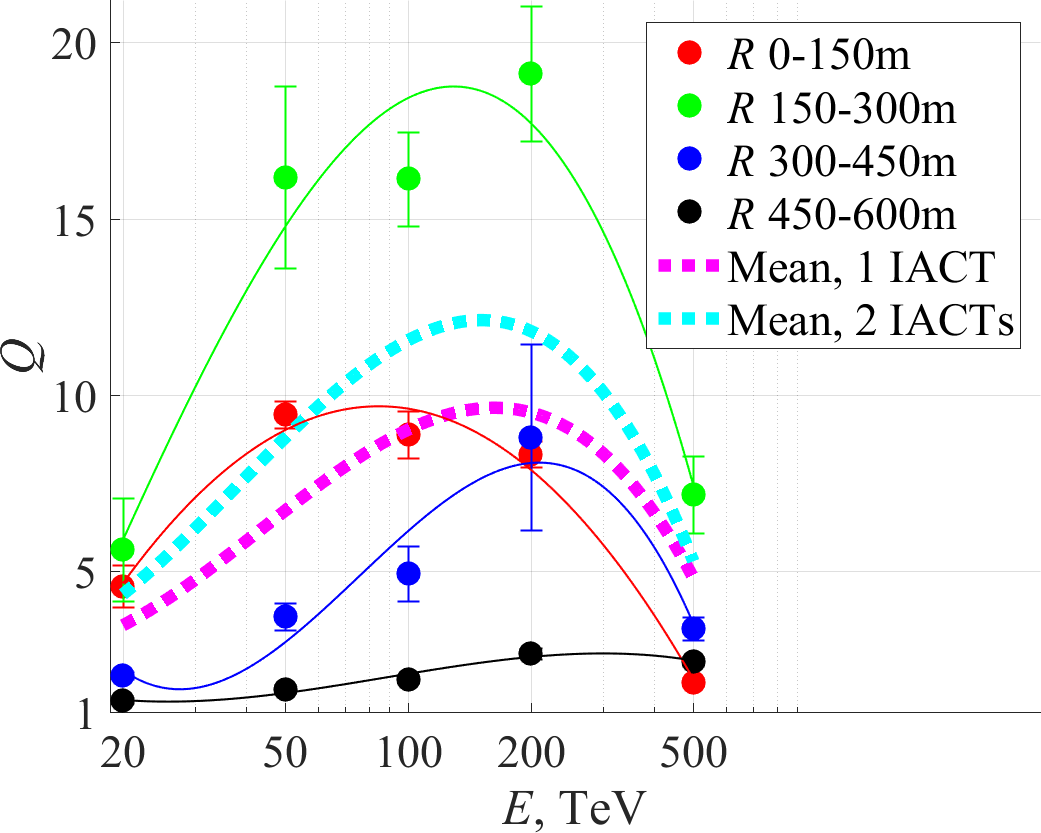}
\caption{Q factor depending on energy and distance. The dashed lines are averaged over the whole installation area with one or two IACTs.}
\label{fig-3}       
\end{figure}

Fig.~\ref{fig-3} demonstrates the results of Q factor computation for a set of energies and core distance intervals. As can be seen, the cosmic ray background suppression about 100 times ($Q$$\sim$5) is to be achieved on a distance up to 450 m for an  energy greater than 50 TeV.

In Fig.~\ref{fig-3} we also present the mean value of $Q(E)$ averaged over the distance between the shower core and the nearest IACT. The upper curve $Q(R)$ represents the variant of two IACTs located at a distance 320~m apart from each other, the lower one -- one IACT. A sensitivity of the gamma-ray observatory depends on the efficiency of background rejection as $\sim$Q$^{-1}(E)$. Therefore, from Fig. \ref{fig-3} the sensitivity improvement due to adding IACTs can be estimated. 

\subsection{$Q$ dependence on core distance}
As follows from the Q factor dependence on the core distance (Fig.~\ref{fig-3}), we can successfully use only one single IACT for efficiently selecting gamma ray showers up to $\sim$450~m in a hybrid installation. This situation differs from a stand-alone IACT, when the distance $\sim$100--150~m is a limit \cite{4}. Therefore, the distance between two or more IACTs (if any) as parts of a hybrid installation can be significantly greater than in a stereoscopic system of IACTs (Section \ref{intro}). In particular, the expected location of the second IACT in TAIGA is supposed to be $\sim$320~m apart from the first one.

\subsection{New parameter}

\begin{figure}
\centering
\includegraphics[height=5cm,clip]{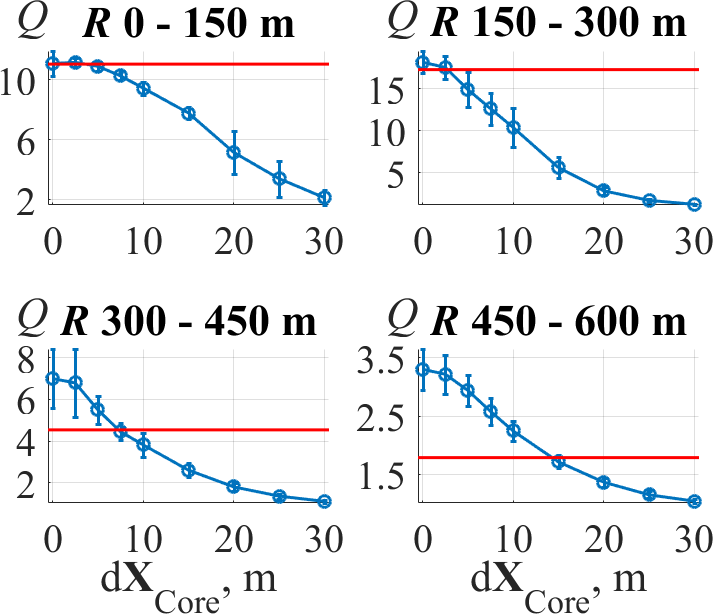}
\caption{Q factor depending on the accuracy of core position determination. Different panels correspond to different distance intervals. Blue points -- using core-azwidth parameter of this work, red line -- using azwidth parameter \cite{11}.}
\label{fig-4}       
\end{figure}

For the case of a core-azwidth parameter the optimal cut depends also on the accuracy of the shower core reconstruction $d\textbf{X}_{Core}$. As a result, the quality factor $Q$ also depends on $d\textbf{X}_{Core}$. In Fig.~\ref{fig-4} we present this dependence for different distance intervals. The meaning of the figure is that the new parameter proved more efficient than the standard parameter azwidth \cite{11} only for a distance greater than $\sim$450~m. This result was obtained given the core position accuracy $\sim$10--15~m. However, the work aiming at the accuracy improvement is now in progress \cite{14}.

\subsection{Possibility to use the IACT without the data from the timing array}

\begin{figure}[!t]
\centering
\includegraphics[height=6.5cm,clip]{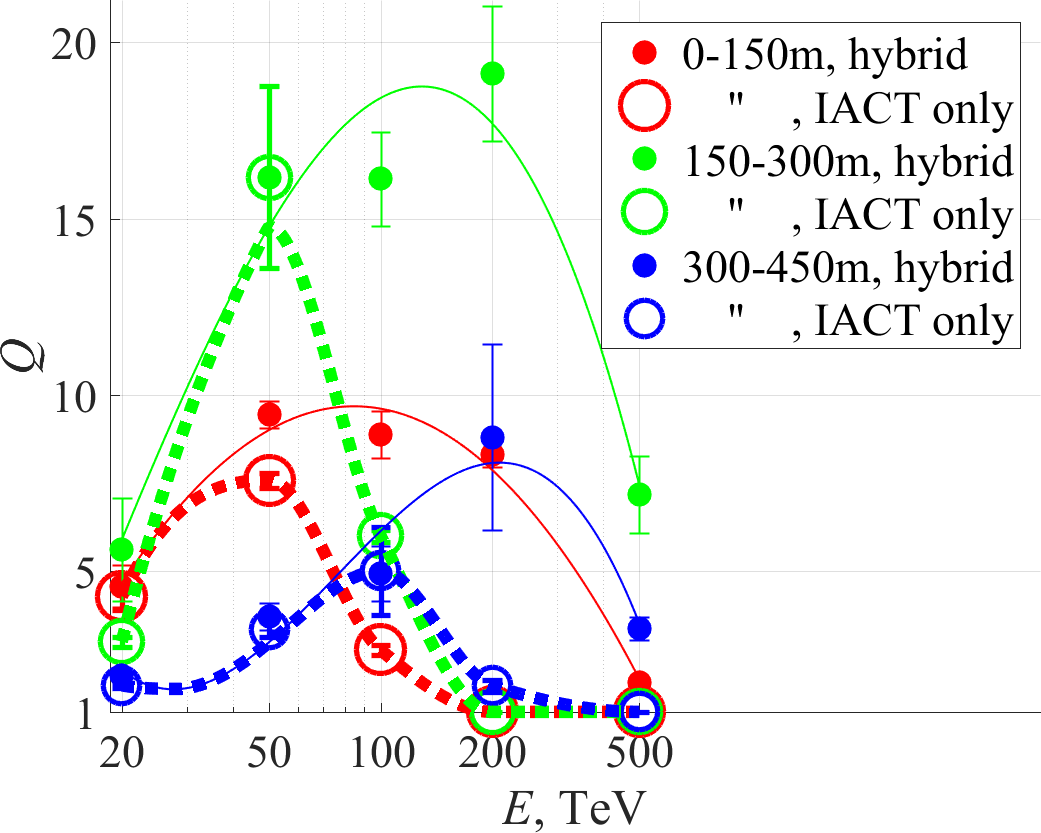}
\caption{Q factor decrease in the absence of timing array data. Solid lines and filled circles: timing array data is present (hybrid technique), dashed lines and open circles: timing array data is absent (stand-alone IACT).}
\label{fig-5}       
\end{figure}

A possibility to use the IACT without the data from the timing array (energy and distance to the core position) was also investigated in the energy interval 20--500~TeV to check the influence of the hybrid technique components on the selection quality. The Q factor obtained for this case is presented in Fig. \ref{fig-5}. The figure demonstrates a major influence of timing array data on the selection quality at high energy.

Therefore, only the hybrid technique can lead to a proper functioning of a single IACT in the conditions of our research (the energy range 20--500 TeV and distances up to 600~m). However, in a hybrid installation this separate use of an IACT could be implemented at low energy (less than $\sim$30~TeV), because in that case the shower core reconstruction and energy determination in a timing array of TAIGA is not available \cite{8}. 

\section{Discussion}
\label{sec-5}

Unlike stereosystems of IACT, a hybrid installation uses images only to select gamma rays, but not to solve other tasks. Therefore, in contrast to standard IACT systems, an image analysis in a hybrid technique can satisfy the only criterion of $Q$ maximization. This helped us to perform full optimization and achieve very high selection quality as well as identify optimal values of cuts for selection and image cleaning different from those of IACT systems.

\section{Conclusions}
\label{sec-6}
We performed a simulation-based methodical study on hybrid reconstruction techniques of both types of Cherenkov detectors: IACT + timing array. As a result we found an optimal set of imaging and timing parameters to be combined, as well as optimal procedures of image cleaning and gamma event selection. The cosmic ray background suppression about 100 times (Q factor $\sim$5) is to be achieved for a distance up to 450~m for  energies greater than 50~TeV. A separate use of the IACT (without the timing array) allows for only a suppression one ninth as weak as that of the joint operation (Q factor value one third as small) at 100 TeV. The investigation confirms in a quantitative way the efficiency of combining the timing-array technique with IACTs in the TAIGA gamma-ray observatory. All the results were obtained for the case that has never been investigated in other experiments. 

The study was supported by the Russian Foundation for Basic Research, project no. 16-29-13035.


\begin{thebibliography}{}

\bibitem{1}
F. Aharonian, \textit {Very High Energy Cosmic Gamma Radiation. A Crucial Window on the Extreme Universe} (World Scientific, 2004)
\bibitem{2}
T.C. Weekes, M.F. Cawley, D.J. Fegan et al., ApJ, \textbf{342}, 379 (1989) 
\bibitem{3}
A. Kohnle, F. Aharonian, A. Akhperdzhanian et al., Astropart. Phys., \textbf{5}, 119 (1996)
\bibitem{4}
A.M. Hillas, Space Sci. Rev., \textbf{6}, 17 (1996)
\bibitem{5}
P. Baillon, L. Behr, S. Danagoulian et al., Astropart. Phys., \textbf{1}, 341 (1993) 
\bibitem{6}
A. Karle, M. Merck, R. Plaga et al., Astropart. Phys., \textbf{3}, 321 (1995) 
\bibitem{7}
A.Sh.M. Elshoukrofy, E.B. Postnikov, L.G. Sveshnikova, J. Phys.: Conf. Ser. (to be published, 2017)
\bibitem{8}
N. Budnev, I. Astapov, P. Bezyazeekov et al., J. Phys.: Conf. Ser., \textbf{718}, 052006 (2016)
\bibitem{9}
M. Kunnas et al., J. Phys.: Conf. Ser., \textbf{632}, 012040 (2015)
\bibitem{10}
D. Heck,  J. Knapp, J.N. Capdevielle et al., \textit{Report FZKA 6019} (Forschungszentrum Karlsruhe, 1998) 
\bibitem{11}
A.M. Hillas, \textit{Proc. 19th ICRC}, \textbf{3}, 445 (NASA Conf. Publ., La Jolla, 1985)
\bibitem{12}
E.B. Postnikov, A.A. Grinyuk, L.A. Kuzmichev et al., Bull. Russ. Acad. Sci. Phys., \textbf{4} (to be published, 2017)
\bibitem{13}
A. Daum et al., Astropart. Phys., \textbf{8}, 1 (1997) 
\bibitem{14}
A.Sh.M. Elshoukrofy, E.B. Postnikov, E.E. Korosteleva et al., these proceedings
\end{thebibliography}
\end{document}